\documentclass[conference]{IEEEtran}

\usepackage[utf8]{inputenc}

\usepackage[
  colorlinks = true,
  linkcolor = blue,
  urlcolor  = blue,
  citecolor = blue,
  anchorcolor = blue]{hyperref}

\usepackage{cite}
\usepackage{xfp}
\usepackage{amsmath,amssymb,amsfonts}
\usepackage{graphicx}
\usepackage{textcomp}
\usepackage{xcolor}
\usepackage{xspace}
\usepackage{ifthen}
\usepackage{wasysym}
\usepackage{color,colortbl}
\usepackage{makecell}
\usepackage{tabularx}
\usepackage{pifont}
\usepackage{booktabs}
\usepackage{tikz}
\usetikzlibrary{shapes.geometric, arrows}
\usepackage{algorithm}
\usepackage{algpseudocode}
\usepackage{array,multirow}
\usepackage{booktabs}
\usepackage{url}
\usepackage{caption}
\usepackage{subcaption}
\usepackage{soul,color}

\DeclareMathOperator*{\argmin}{arg\,min}

\graphicspath{{figures/}}
\newcommand{\omnifiguresscalingfactor}{0.485}

\definecolor{__red}{rgb}{0.8,0.1,0.1}
\definecolor{__green}{rgb}{0.1,0.6,0.1}
\definecolor{verylightgray}{gray}{0.95}

\makeatletter
\g@addto@macro{\endtabular}{\rowfont{}}%
\makeatother
\newcommand{\rowfonttype}{}%
\newcommand{\rowfont}[1]{%
   \gdef\rowfonttype{#1}#1%
}

\newsavebox\CBox

\newboolean{showcomments}
\setboolean{showcomments}{false}
\ifthenelse{\boolean{showcomments}}
{ \newcommand{\mynote}[3]{
    \fbox{\bfseries\sffamily\scriptsize#1}
    {\small$\blacktriangleright$\textsf{\emph{\color{#3}{#2}}}$\blacktriangleleft$}}}
{ \newcommand{\mynote}[3]{}}

\newcommand{\sysname}{{StreamBed}\xspace}
\newcommand{\flink}{Flink\xspace}
\newcommand{\flinklong}{Apache Flink\xspace}

\begin{document}

\title{StreamBed: capacity planning for stream processing}

\author{\IEEEauthorblockN{Guillaume Rosinosky}
\IEEEauthorblockA{UCLouvain, Belgium\\
guillaume.rosinosky@uclouvain.be}
\and
\IEEEauthorblockN{Donatien Schmitz}
\IEEEauthorblockA{UCLouvain, Belgium \\
donatien.schmitz@uclouvain.be}
\and
\IEEEauthorblockN{Etienne Rivière}
\IEEEauthorblockA{UCLouvain, Belgium \\
etienne.riviere@uclouvain.be}
}

\maketitle

\begin{abstract}

  \sysname is a capacity planning system for stream processing.%
  It predicts, ahead of any production deployment, the resources that a query will require to process an incoming data rate sustainably, and the appropriate configuration of these resources.
  \sysname builds a capacity planning model by piloting a series of runs of the target query in a small-scale, controlled testbed.
  We implement \sysname for the popular \flink DSP engine.
  Our evaluation with large-scale queries of the Nexmark benchmark demonstrates that \sysname can effectively and accurately predict capacity requirements for jobs spanning more than 1,000 cores using a testbed of only 48 cores.
\end{abstract}

\begin{IEEEkeywords}
capacity planning, stream processing, Flink
\end{IEEEkeywords}

\section{Introduction}
\label{sec:introduction}

Distributed Stream Processing (DSP) allows the processing of high-velocity data streams.
DSP is a key component of analytics and decision-support systems, extracting information of interest from continuous data with low latency.
Twitter~\cite{templeton_twitter_2022} and Uber~\cite{fu2021real} report their use of DSP for processing trillion of events per day, for volumes in the order of hundreds of petabytes.
DSP is also often a key component in data lake platforms~\cite{hai2023data}, allowing data scientists and analysts to deploy continuous queries on-demand, as exemplified by King~\cite{fora_king_2016}, an online game operator processing up to 30 billion events daily. %

\begin{figure}[t!]
    \centering \includegraphics[scale=\omnifiguresscalingfactor]{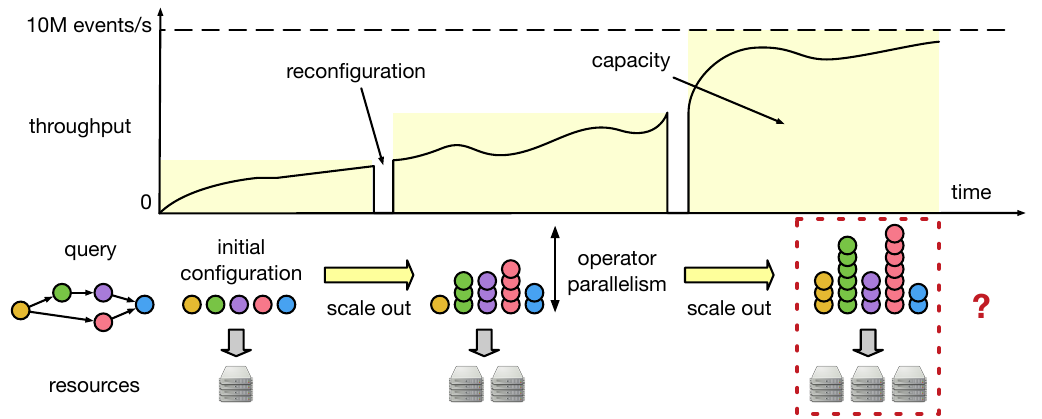}
    \vspace{-1mm}
    \caption{DSP engines scale out to handle increasing throughputs.
    Reconfigurations apply the necessary parallelism for a query operators (its \emph{configuration}) to increase its capacity.
    Determining \emph{in advance} the necessary resources budget and its configuration for a target peak capacity is our objective.
    \vspace{-3mm}}
    \label{fig:intro_elastic_scaling}
\end{figure}

A number of DSP engines support the efficient execution of stream processing queries, e.g.,
  Twitter Heron~\cite{kulkarni2015twitter},
  Google Cloud Dataflow~\cite{akidau2015dataflow}, or
  Spark Streaming~\cite{zaharia2013discretized}.
\flinklong~\cite{carbone2015apache} is one of the most active open-source projects in this space.
Similarly to other engines, \flink leverages massively parallel processing of incoming data.
In \flink, a user query is expressed as a graph of processing \emph{operators}, and each operator is supported by multiple \emph{tasks}.
\flink, like other engines~\cite{roger2019comprehensive,cardellini2022runtime}, offers \emph{elastic} scaling as illustrated by Figure~\ref{fig:intro_elastic_scaling}.
When the level of parallelism of operators (i.e., the query \emph{configuration}) is insufficient, elastic scaling policies trigger a reconfiguration to add new tasks, if necessary requesting new resources from a resource management system such as Kubernetes~\cite{flink_integration_of_ds2_k8s_autoscaler}.
A reconfiguration typically requires snapshotting and restoring the state of the query, incurring downtime~\cite{carbone2017state,barazzutti2014elastic,castro2013integrating,gu2022meces}.

\smallskip
\noindent
\textbf{Motivation.}
It is hard to know, before the full-scale deployment of a new query, the resources budget that it will require, or how these resources will have to be configured to \emph{sustainably} ingest a given, target workload (i.e., achieve a given \emph{capacity}).
For instance, the use of a total of twelve servers and the parallelism of each of the five operators in Figure~\ref{fig:intro_elastic_scaling}'s configuration at peak capacity is the result of runtime decisions that depend on the scaling behaviors of operators and their reaction to operational conditions, e.g., the amount of RAM provided to each of the tasks, the performance of the storage sub-system, or characteristics of the input stream.
The scaling behavior of queries is often non-linear, disallowing simple proportionality-based predictions, and they often exhibit non-trivial resource usage and performance patterns, e.g., load spikes due to straggler events~\cite{farhat2020leaving} or skew between tasks.

The uncertainty in the resource requirements of DSP jobs often leads practitioners to exercise caution and over-provision their support infrastructure.
An infrastructure that cannot scale out to the needs of a query will result in cascading failures and query termination~\cite{flink_forward_autopilot}.
When DSP jobs are deployed over a public cloud, the problem becomes that of mastering costs, as uncontrolled scale-out may lead to paying for a large amount of on-demand resources.

\begin{figure}[t]
    \centering
    \includegraphics[scale=\omnifiguresscalingfactor]{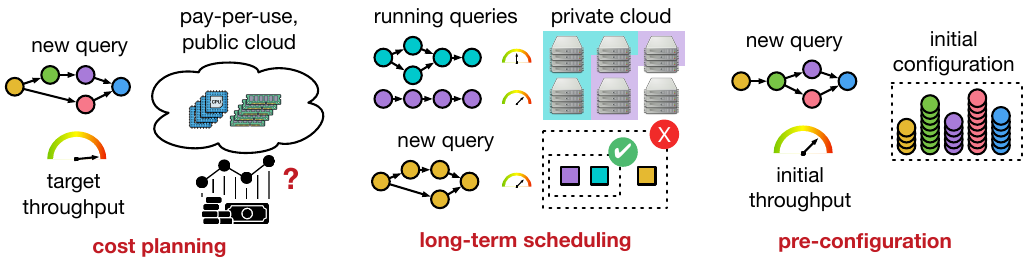}
    \vspace{-1mm}
    \caption{
    \sysname enables capacity planning for stream processing.
    On the left, it helps to determine the cost of running a large query at a certain throughput and for a certain duration in the cloud.
    In the middle, it allows for determining if a new query can be safely deployed alongside others in a private cloud with limited resources.
    On the right, it returns an initial configuration for a query, avoiding the costs and service interruptions of many scale-out reconfigurations.
    \vspace{-3mm}}
    \label{fig:intro_uses}
\end{figure}

\smallskip
\noindent
\textbf{Contributions.}
We present \sysname, a capacity planning system for DSP.
\sysname predicts, ahead of any production deployment, the resources budget and profile for supporting sustainably a specific query, and their configuration.
\sysname target use cases as illustrated by Figure~\ref{fig:intro_uses}.

Previous work proposed benchmarking strategies~\cite{chintapalli2016benchmarking,lopez2016performance,henning2021theodolite,henning2022benchmark,li2016performance,kalim2019caladrius}, requiring to deploy queries at a production scale and incurring significant costs.
Techniques based on models of a specific query, e.g., using queuing theory~\cite{truong2017predicting,truong2018performance} or Markovian processes~\cite{russo2021mead}, or building a \emph{general} model of queries characteristics and their link to performance~\cite{khoshkbarforoushha2015resource, khoshkbarforoushha2016distribution,heinrich2022zero,agnihotri2023zero} do not require large-scale deployments but often fail to capture the complex scaling behavior of a DSP engine used close to its full capacity, and the impact of the characteristics of input data.
In contrast with these approaches, \sysname analyzes the behavior of a submitted query running in a controlled, small-scale testbed under different configurations.
The size of the testbed used for capacity planning is much smaller than the size of a production system, with typically one to two orders of magnitude less CPU cores and memory. 
\sysname builds a \emph{capacity planning model} based on these controlled runs using actual data, carefully exploring different resource budgets and identifying for each budget its optimal configuration and capacity. 

We implement \sysname for \flinklong as a combination of three nested mechanisms.
A \emph{Resource Explorer} drives the construction of the capacity planning model, using Bayesian Optimization to explore different resource budgets and different allocation granularities (RAM and CPU).
For a given resources budget, it is necessary to identify its maximum \emph{sustainable} throughput (MST)~\cite{karimov2018benchmarking, imai2017maximum, chu2020maximum}, i.e., the volume of data that can be ingested without provoking instabilities and, eventually, crashes.
The MST of a query depends on its configuration, i.e., the distribution of available resources to the different operators. %
A \emph{Configuration Optimizer} determines the best possible configuration for a given resource budget by extending the state-of-the-art, reactive elastic scaling algorithm DS2~\cite{kalavri2018three} for a predictive context and bounded resources.
The Configuration Optimizer needs to collect information from actual, small-scale runs of the target job.
Determining the MST of a configuration is difficult due to buffering and ramp-up effects, instabilities at rates higher than what is sustainable, and the need to make this measurement as quickly as possible.
The \emph{Capacity Estimator} overcomes these challenges.
Using controlled data injection in bursts of varying rates, it can reliably determine the MST of a configuration. %
Based on the MST estimations for varying amounts of resources, the \emph{Resource Explorer} uses regressions to build a capacity planning model, serving as a fine-grain query scaling and configuration oracle.

We evaluate \sysname in a cluster of 85 nodes, for a total of 1,344 cores and 7.5~TB of RAM.
Out of these, \sysname uses 48 cores and 192~GB of RAM for controlled runs; the rest supports production deployments and data lake services.
We use 5 representative queries from the Nexmark benchmarking suite~\cite{tucker2008nexmark}.
Our results show that \sysname can accurately predict the volume of necessary resources with a low cores.hours budget, for queries reaching more than 1,000 cores or ingesting up to 190 million events per second.
Our evaluation of \sysname predictions at a production scale shows that the system avoids over- or under-provisioning and derives configuration that can sustainably inject the targeted loads, even for complex, stateful queries.

\smallskip
\noindent
\textbf{Outline.}
We present the background on \flink and DSP as well as state-of-the-art elastic scaling mechanisms in Section~\ref{sec:background}.
We detail our operational assumptions and give an overview of \sysname components in Section~\ref{sec:overview}, before detailing the Capacity Estimator (Section~\ref{sec:capacity_estimator}), the Configuration Optimizer (Section~\ref{sec:configuration_optimizer}), and the Resource Explorer (Section~\ref{sec:resource_explorer}).
We discuss implementation details in Section~\ref{sec:implementation}, and present our evaluation of \sysname in Section~\ref{sec:evaluation}.
We follow with related work in Section~\ref{sec:related_work} and a conclusion in Section~\ref{sec:conclusion}.

\section{Background}
\label{sec:background}

We provide background information necessary to introduce our contributions, namely the principles of Distributed Stream Processing with a focus on \flinklong and state-of-the-art techniques for elastically scaling such systems.

\begin{figure}[t]
    \centering
    \includegraphics[scale=\omnifiguresscalingfactor]{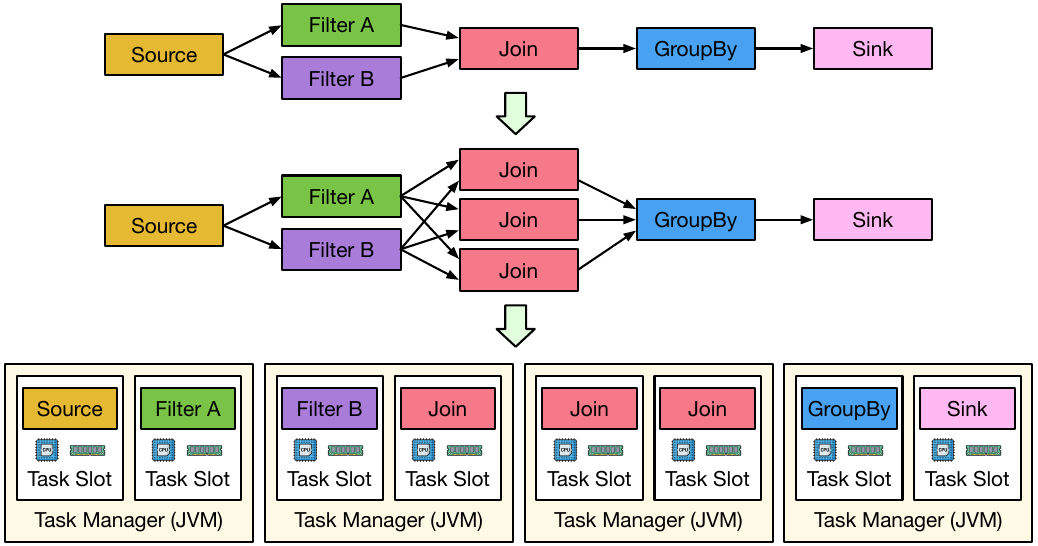}
    \vspace{-1mm}
    \caption{A \flink job with 4 operators, a source, and a sink (top).
    The Join operator is scaled-out to a parallelism of three (middle).
    The resulting eight tasks are distributed to available task slots on four task managers (bottom).
    \vspace{-3mm}}
    \label{fig:background_example}
\end{figure}

\smallskip
\noindent
\textbf{\flinklong.}
A DSP engine supports queries over continuous data, or \emph{jobs}, implemented as an oriented graph of interconnected \emph{operators} as illustrated by the top-most example in Figure~\ref{fig:background_example}.
Each operator implements a basic operation on the data flow(s) it receives as its input, and outputs a stream of events to be consumed by downstream operators. %
\flink supports a variety of operators, selected from a standard library or defined by programmers.
It is also increasingly common that the operator graph is generated from SQL queries~\cite{rabl2016apache,begoli2019one}.%
While some operators are stateless (e.g., map or filter), others may need to persist state across events (e.g., join or group by) such as data structures over a window of time or events.
Sources and Sinks are specific operators injecting, respectively outputting data from and to the external world.

Deployment of a \flink job is under the responsibility of a centralized Job Manager node orchestrating several Task Managers (TM), represented at the bottom of Figure~\ref{fig:background_example}. %
Each TM runs in its own process (JVM) and offers a number of Task Slots (TS).
\flink, similarly to other DSP engines, assumes homogeneous task slots, each assigned to one CPU core but with a configurable amount of memory forming the task \emph{profile}. %
In Figure~\ref{fig:background_example}, the parallelism of operators is 1, except for the Join using a parallelism of 3.
The resulting 8 tasks are dispatched by the Job Manager to the available 4 TMs.
The flow of events to a parallelized operator, e.g., from Filter A to the Join, is partitioned based on the key associated with events~\cite{carbone2017state}.
Each task maintains a buffer of incoming events.
\emph{Back-pressure} mechanisms regulate the production of events: a task can instruct its upstream operator(s) to slow down production when its buffer goes over a certain fill rate.

In production, stateful \flink operators use RocksDB~\cite{dong2021rocksdb}, a key-value store based on a Log-Structured-Merge tree.
RocksDB is tailored for efficient write operations and optimizes wear and performance on SSDs.
RocksDB combines in-memory and SSD storage, the former playing the role of a cache.
Insufficient memory may lead to an increase in access to the SSD and a drop in performance, making memory an important factor in performance for stateful operators. 

\smallskip
\noindent
\textbf{Elastic Scaling.}
Elastic scaling allows coping with varying input rates for a DSP job, as imposed by its data sources.
A configuration maps each operator to a level of parallelism, i.e., the number of tasks that supports it.
A reconfiguration adapts this parallelism and is generally triggered by indicators such as the amount of back-pressure or system-level metrics such as the average CPU consumption of TMs~\cite{carbone2017state}.
R{\"o}ger and Mayer~\cite{roger2019comprehensive} and Carellini \emph{et al.}~\cite{cardellini2022runtime} present comprehensive surveys of elastic scaling approaches for DSP.
In what follows, we focus on the current state of the art, the DS2 algorithm of Kalavri \emph{et al.}~\cite{kalavri2018three}.
DS2 is used by Linkedin~\cite{singh2020auto} and was recently integrated with Flink~\cite{flink_integration_of_ds2} as the elastic scaling mechanism for its Kubernetes-based resource manager~\cite{flink_integration_of_ds2_k8s_autoscaler}.

\begin{figure}[t]
    \centering
    \includegraphics[scale=\omnifiguresscalingfactor]{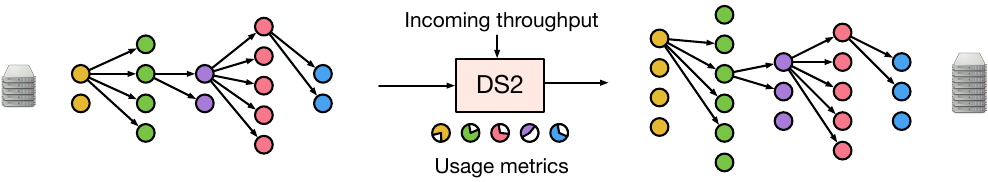}
    \vspace{-1mm}
    \caption{Illustration of a reconfiguration using DS2~\cite{kalavri2018three}.
    Only a subset of the edges is shown for clarity.
    \vspace{-3mm}}
    \label{fig:ds2}
\end{figure}

The key idea underlying DS2, illustrated by Figure~\ref{fig:ds2}, is to determine simultaneously the level of parallelism for all operators of a job undergoing a scale-out rather than scaling out each operator independently~\cite{roger2019comprehensive,cardellini2022runtime}.
DS2 collects rate measurements for all operators' tasks in the DSP job, and in particular a measure of their \emph{busyness}, i.e., the ratio of time spent actually processing incoming events.
DS2 then determines the \emph{true processing rate} of operators, assuming linear scaling of their capacity, and applies a one-pass computation to determine the level of parallelism of operators required for the current workload. %
DS2 has limitations that prevent from using it for capacity planning on the basis of a small-scale run.
DS2 is based on optimistic, linear scaling assumptions for operators.
While this assumption is valid in certain cases, additional steps of rate measurement and optimizations are often necessary to account for non-uniform scaling profiles, requiring up to two additional reconfiguration steps for complex queries~\cite{kalavri2018three}.
These additional steps are performed on the scaled-out job running in production, in contrast with our capacity planning scenario where the estimation must only rely on pre-production, small-scale runs.
The deviation from a linear scaling model can have different origins.
These include first the impact of memory, which is not considered by DS2.
Memory pressure and the resulting performance of the storage subsystem lead to slowdowns that may appear only past a certain state size.
Second, the distribution of keys in events exchanged between two operators is typically not uniform but exhibits some skew: linear scaling assumption may result in overloaded and underloaded tasks, exacerbated by increases in scale.
Third, the use of windowed operations in queries, and more generally operators with bursts in processing time for a subset of events, lead to a phenomenon of \emph{stragglers} events~\cite{farhat2020leaving} that pile up, and are sent and consumed in bursts.
Stragglers require provisioning operators to handle the associated load peaks, an aspect that DS2 does not anticipate.

\section{Overview}
\label{sec:overview}

We detail \sysname objectives, operational model, and target workloads.
Then, we provide an overview of its components and their interactions. %

\smallskip
\noindent
\textbf{Objectives and assumptions.}
\sysname targets long-run\-ning queries that need to scale out to hundreds of cores for their execution.
We consider a system formed of two clusters: a large one is dedicated to production deployments, while a much smaller one is dedicated to capacity planning.
The two clusters use the same hardware and network interconnect, to make observations on one transposable onto the other.

The goal of \sysname is to build a capacity planning model for a \emph{specific} DSP query provided by the user.
We assume that a dataset, representative of the input of the job, is also provided by that user, e.g., using historical data for the corresponding source stored in a data lake.
We do not require, however, that the query ran previously on this dataset or any other dataset.
The query does not have to be modified by the user, but \sysname needs to be informed of fields in events' schemas used to represent event time, if any.
This information is needed to replace recorded, historical times with emulated times in controlled runs of the query.

We assume that the submitted DSP job uses an arbitrary combination of stateless and \emph{windowed} operators including joins, as is common for continuous queries~\cite{verwiebe2023survey}.
This means that operators do not accumulate state without limit, but operate on a working set formed of the events received over a window, defined based on time and/or a number of events.\footnote{We do not target \emph{batch} processing jobs~\cite{carbone2015apache}, where the processing throughput is dictated by the job deployed capacity and not by its sources, and where some operators such as non-windowed joins may keep accumulating state for all events of an input before they can process another.}%
We do not make prior assumptions on the length of this window or the sizes of operators' working sets.

\sysname builds a model allowing querying the necessary resources budget for different input rates (i.e., the number of task slots and their resource profile), as well as their appropriate configuration (parallelism for each operator).
The returned configurations must be able to sustain the requested throughput and avoid over- or under-provisioning.\footnote{We emphasize that \sysname capacity planning is not meant to replace runtime, elastic scaling mechanisms, but rather to complement them.}

\sysname handles skew (i.e., the non-uniform distribution of event keys for an operator, and the resulting imbalance in its tasks load) and the unpredictable load peaks due to stragglers by considering the actual maximal sustainable throughput (MST) of the query under a specific resources budget, and the evolution of this MST with varying budgets.%
In contrast with DS2~\cite{kalavri2018three}, this allows for predicting the peak capacity that a configuration will yield despite these unbalances.%

\smallskip
\noindent
\textbf{The case of Source and Sink operators.}
\sysname builds a capacity planning model for all operators except Sources.
The motivation for this choice is that Source operators are diverse (e.g., some may implement format translation or decompression) but are stateless and scale linearly.
Moreover, the same Source operators are likely to be reused across multiple jobs, allowing for the collection of a general profile of their capacity and performing capacity planning separately.
We include, however, simple ``blackhole'' Sink operators to determine the volume of final events received from upstream operators.
The necessary parallelism of Sink operators used in production will depend on their nature, e.g., pushing events to Kafka will have a different cost than spilling them to disk after compression.
As for the sources, the scaling of Sink operators is linear, so this parallelism can be determined based on the volume of events received by the blackhole Sink.
We finally consider queries with a single source operator, possibly ingesting events of different types.

\begin{figure}[t!]
    \centering
    \includegraphics[scale=\omnifiguresscalingfactor]{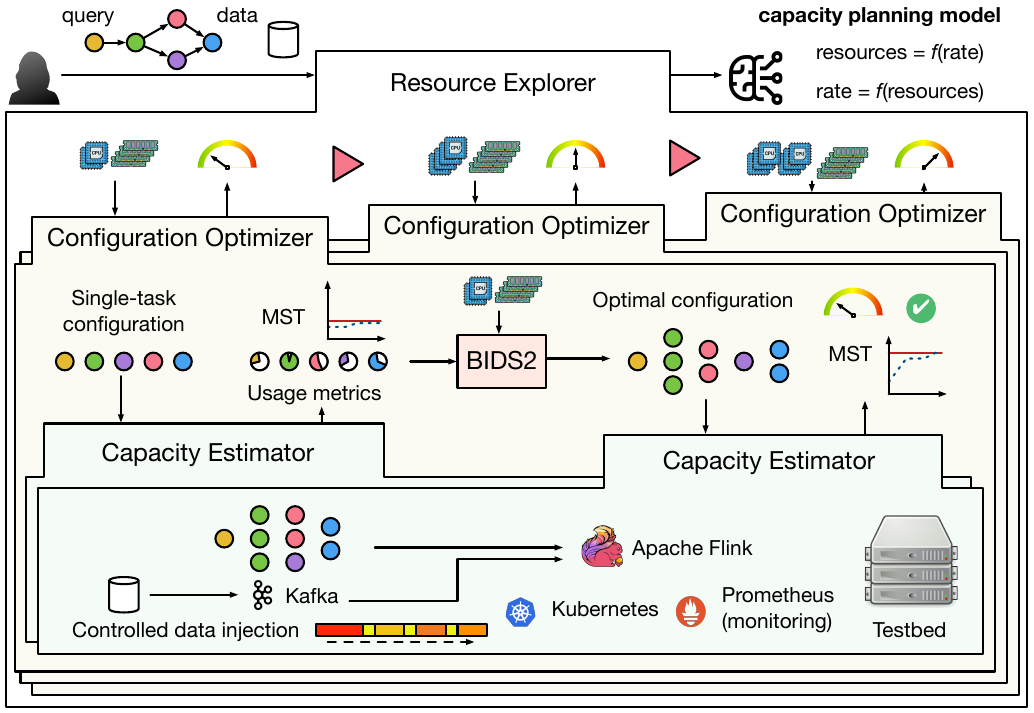}
    \caption{\sysname workflow. 
    The user submits her query and a representative dataset.
    The \textbf{resource explorer} builds a capacity planning model from capacity estimations with small resource budgets.
    For each budget, the \textbf{configuration optimizer} determines the best resource allocation and its processing capacity (max sustainable throughput, MST) via controlled benchmarking by the \textbf{capacity estimator}.
    \vspace{-4mm}}
    \label{fig:overview}
\end{figure}

\smallskip
\noindent
\textbf{\sysname components.}
The general workflow and the three nested components of \sysname are illustrated in Figure~\ref{fig:overview}.
At the top-most level, the \textbf{Resource Explorer (RE)} drives the exploration of different \emph{resource budgets}, i.e., numbers of task slots with a given resource profile.
For each budget, the RE collects the achievable capacity or Maximal Sustainable Throughput (MST)~\cite{karimov2018benchmarking, imai2017maximum, chu2020maximum} and the associated configuration, which it uses to build a capacity planning model.
The choice of profiles is driven by a Bayesian Optimization algorithm for a set of target candidate models. %
In Figure~\ref{fig:overview}, the RE evaluates three such resource budgets.

The determination of the best configuration and of MST for a given resource budget is under the responsibility of the \textbf{Configuration Optimizer (CO)}.
For resource profiles that have not yet been evaluated, a first step evaluates in-situ (using the lowest-level component, the Capacity Estimator) a minimal configuration with a parallelism of 1 for each operator, to collect usage metrics.
Then, the CO leverages BIDS2 (Bounded-Inverse DS2), our evolution of the DS2 algorithm able to determine in one pass and for a bounded resource budget, the best possible configuration.
In Figure~\ref{fig:overview}, the CO determines the MST for the smallest of the budgets submitted by the RE for evaluation.

The determination of the MST of a given configuration is performed by the \textbf{Capacity Estimator (CE)}, which uses in-situ, controlled runs of the job in the small-scale testbed.
Determining the MST experimentally is a complex task.
Performance tends to fluctuate heavily during \emph{ramp-up} phases, while the state of operators and buffers between tasks are not stabilized at their peak size.
Performance is also generally highly unstable when operating above the MST, and queries may exhibit important variations of resource usage over time.
We devise a controlled data injection strategy that can effectively determine the MST of a configuration by stress-testing it and adjusting input rates using a dichotomous strategy.

In the following sections, we detail these components in a bottom-up fashion, starting with the Capacity Estimator (\S\ref{sec:capacity_estimator}), following with the Configuration Optimizer (\S\ref{sec:configuration_optimizer}), and finishing with the Resource Explorer (\S\ref{sec:resource_explorer}).

\section{Capacity Estimator}
\label{sec:capacity_estimator}

The role of the Capacity Estimator (CE hereafter) is to determine experimentally the Maximal Achievable Throughput, or MST, of a job in a specific configuration, upon request of the Configuration Optimizer.
The MST, a common performance metric for DSP~\cite{karimov2018benchmarking, imai2017maximum, chu2020maximum}, is the highest possible average actual rate that can be processed by the configuration, that corresponds to the injected rate (i.e., no piling up of unprocessed records between the Source operator and the rest of the job).%
The goal of the CE is to estimate the MST in a small amount of time, while guaranteeing its accuracy, in particular with respect to the impact of stateful operators, skew, and stragglers.
The CE must also take into account that injected rates higher than the MST often lead to chaotic behaviors such as unsteady actual processing rates.

\smallskip
\noindent
\textbf{Load injection.}
The CE uses controlled load injection, attaching rate-limited sources to the job.
These sources can connect with a distributed storage in a data lake, allowing the use of data at rest as a representative input.
Alternatively, \sysname can employ as a source an online pseudo-random event generator provided by the user.
In both cases, sources attempt to inject events at up to a fixed target rate but abide by the back-pressure received from downstream operators. 

\smallskip
\noindent
\textbf{Time-based operators.}
If the query contains time-based operations, such as time-windowed aggregations or joins, it is necessary to replace event times in the data at rest with a time matching the target rate upon replay.
\sysname automatically replaces fields of the events schema declared by the user as representing event time with the \texttt{proctime()} \flink dynamic call~\cite{flink_time_attributes}, allowing time-based operators to consider their processing time instead.

\smallskip
\noindent
\textbf{Warmup phase.}
We start by observing that a newly-started DSP job will typically accept incoming load at a much higher throughput than its steady-state MST.
Two factors explain this.
First, each edge of the job graph is associated with buffers; back-pressure mechanisms only kick in when the fill rates of these buffers reach a threshold.
Initially, empty buffers lead to a temporarily higher ``absorption capacity'' for the job.
Second, stateful operators start with an empty state that gradually grows until it reaches a steady state (e.g., a full sliding window of events).%
Initially, operations requiring to access state can be much faster, possibly biasing measurements.
These observations lead us to the need for a \emph{warmup} phase prior to any measurement, to observe performance metrics on a stabilized job.

\begin{figure}[t!]
    \centering
    \includegraphics[width=1\linewidth]{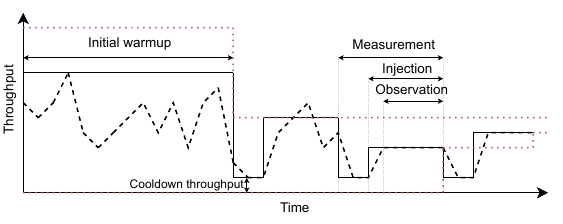}
    \vspace{-7mm}
    \caption{The Capacity Estimator uses a evaluation strategy based on fixed-rate load injection, with a dichotomous search following an initial warmup phase.
    Black plain lines show the target input rate, while black dashed lines shows the actual, measured processing rate.
    Dashed, thin red lines present the maximal and minimal rates $\text{max}_r$ and $\text{min}_r$.
    \vspace{-3mm}
    }
    \label{fig:mst_estimator}
\end{figure}

\smallskip
\noindent
\textbf{Evaluation strategy.}
A policy decides on variations of the input rate, and on when to measure the actual processing rate.

A naive approach to determining the MST would be to start from a low target rate and gradually increase it until the source starts observing a certain back-pressure level.
This approach is inappropriate for two reasons.
First, it mixes the necessary warmup phase with the actual measurement phase, risking estimating the MST of a job that has not reached its steady state.
Second, it does not take into account the inertia of the job under test: A change in input rate leads to cascading effects on buffer occupancies and back-pressure levels.
These effects often take several seconds, and sometimes several minutes, to stabilize.
Using an ever-increasing rate bears risks that this stabilization never occurs.

We propose a strategy, illustrated by Figure~\ref{fig:mst_estimator}, that takes into account these concerns by using fixed target throughputs, including necessary stabilization phases but avoiding measurements during these phases.
It starts with a sufficiently-long warmup phase where we inject throughput at the maximal possible rate, building up sufficient state at stateful operators and filling buffers.
Reaching this steady state can require several minutes of warmup, primarily depending on available memory and the size of the working state of stateful operators.

The warmup phase is followed by the evaluation itself, using a dichotomous approach similar to a binary search.
We consider a \emph{maximal} and a \emph{minimal} rate, $\text{max}_r$ and $\text{min}_r$, initially set to $\infty$ and $0$. 
The search progresses in phases.
In each phase, a target rate $r$ is tested.
If the test is successful, which the algorithm determines as the actual processing rate being equal or very close (i.e., $\geq$99\%\footnote{Due to the use of a rate limiter, the actual rate can never exceed 100\%.}) to the incoming rate then $\text{min}_r=r$.
Otherwise, $\text{max}_r=r$.
The next target is the average between these two values, i.e., $r=(\text{max}_r + \text{min}_r)/2$.
The search stops when reaching a configurable sensibility (e.g., the next value of $r$ is within 1\% of the previous) or after a maximal number of iterations. 

A measurement for a target rate $r$ starts with a \emph{cooldown} phase, using a low (but non-zero) rate allowing operators to process events in their input buffers and, possibly, to recover from their saturation in the previous measurement.
Then, the target rate $r$ is applied for an \emph{injection} phase.
We do not measure the actual processing rate immediately but only after a short duration, to enable a local ramp-up for the new injection.
During the \emph{observation} phase, we observe the achieved rate from the point of view of the source and its stability and collect busyness metrics and actual input rates for all operators.

The metrics for the final measurement are returned along with the MST to the Configuration Optimizer.

\section{Configuration Optimizer}
\label{sec:configuration_optimizer}

The second component of \sysname is the Configuration Optimizer, or CO for short.
The CO receives from the Resource Explorer a query and a resource budget, i.e., a number of task slots with a given profile (amount of RAM).
Its goal is to return the optimal configuration fitting this budget, together with its MST as measured \emph{in situ} using the CE.

The CO requires usage metrics for the execution of a \emph{minimal} configuration, i.e., where each operator has a parallelism of 1.
These usage metrics include the \textit{actual input rate} and the level of \textit{busyness} of each single-task operator, i.e., the percentage of time it spends actually processing events.
The CO maintains a cache of these single-task configuration metrics for different resource profiles, reusing previous measurements when possible. 
When no data exists, the CO requests the CE to evaluate this single-task configuration with the target profile.
An exception to this reuse rule is when the Resource Explorer explicitly requests the evaluation of a single-task configuration as part of its exploration. %

\begin{figure}[t]
    \centering
    \includegraphics[scale=\omnifiguresscalingfactor]{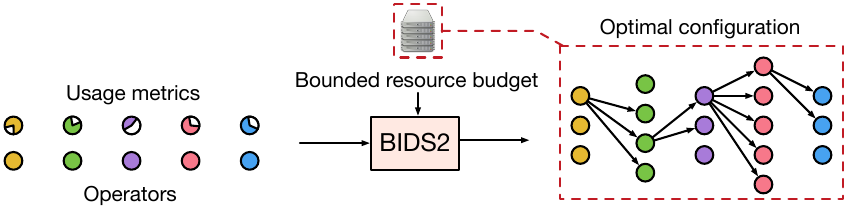}
    \vspace{-1mm}
    \caption{The BIDS2 algorithm determines, from usage metrics collected from a minimal run with a parallelism of 1 for all operators and a bounded resource budget, the optimal configuration fitting that budget.\vspace{-3mm}}
    \label{fig:bids2}
\end{figure}

Based on observed metrics on the single-task configuration, we formulate the problem of determining the \emph{optimal} capacity for the target resource budget, i.e., the highest possible source input rate.
The corresponding \emph{best possible} configuration is the one where the busyness metric of all tasks, across all operators, is the highest possible while avoiding the saturation of any task.
We solve this optimization problem using an algorithm we name BIDS2, for Bounded-Inverse DS2.
In contrast with the original DS2~\cite{kalavri2018three}, BIDS2 does not determine the amount of necessary resources for a given input rate, but determines the optimal configuration for a \emph{bounded} resource budget, as illustrated by Figure~\ref{fig:bids2}.
The configuration resulting from the optimization is given as an input to a call to the CE to determine its MST and experienced busyness levels. %

\begin{table}[b]
  \begin{center}
    \newcolumntype{G}{>{\raggedright\arraybackslash} m{0.105\linewidth} }
    \newcolumntype{H}{>{\raggedright\arraybackslash} m{0.77\linewidth} }
    \renewcommand{\arraystretch}{1.2}
    
    \begin{tabular}{GH}
    \toprule
    
    \textbf{Symbol} & \textbf{Description} \\ 
    \midrule
    $G$ & Query logical graph \\ %
    $\mathcal{P}$ & Task slots budget \\ 
    $o^i$ & Observed true processing rate of a task of operator $i$\\
    $r^i$ & Observed ratio of operator $i$'s input rate over source's rate \\
    \midrule
    $\pi^i$ & Optimal parallelism of the operator $i$ \\
    $\lambda_\mathrm{src}$ & Optimal input rate of the source operator \\
    $\lambda^i$ & Optimal input rate of one operator instance $i$ \\    
    \bottomrule
    
    \end{tabular}
    \caption{Notations used by the BIDS2 algorithm.}
    \label{table:notation}
    \end{center}
\end{table}

\smallskip
\noindent
\textbf{BIDS2 algorithm.}
Table~\ref{table:notation} lists the variables used in the optimization problem.
Our goal is to maximize the throughput of the source, as modeled by Equation~\ref{eq:objective}.

\vspace{-2mm}
\begin{equation}
    \max \lambda_{\mathrm{src}}
    \label{eq:objective}
\end{equation}

To this end, we must find $\pi^i$ the level of parallelism for each operator $i$. 
$\pi^i$ is the decision variable for each operator $i$, $i$ varying between $1$ and the number of operators of the job excluding the sources.
As Kalavri \emph{et al.}~\cite{kalavri2018three}, we assume in this optimization problem a linear relation between the processing rate and the level of busyness: we compute the true processing rate $o^i$ of a task of operator $i$ by dividing its actual processing rate by the average busyness level of its tasks.
While this relation can be considered as linear at a small scale and is sufficient in this module, linearity is not guaranteed at higher scales: in \sysname, we manage this non-linearity when building the model at the Resource Explorer level, as it will be reflected in the different capacities measured for different budgets and profiles.

We compute the optimum processing rate ${\lambda^i}$ of an operator $i$ based on its parallelism $\pi^i$, as given by Equation~\ref{eq:constraint_rate_operators}. 

\vspace{-2mm}
\begin{equation}
    \forall{i}: \; \lambda^i = \pi^i o^i
    \label{eq:constraint_rate_operators}
\end{equation}

Then, we compute using Equation~\ref{eq:constraint_input_rate} the rate of each operator expressed as a function of the decision variable input rate ${\lambda_{src}}$: The ratio $r_i$, precomputed using the actual rates returned by the CE is used as a multiplier to estimate this proportion.

\vspace{-2mm}
\begin{equation}
    \forall{i}:  \lambda_{\mathrm{src}} \cdot r^{i}  \le \lambda^i    
    \label{eq:constraint_input_rate}
\end{equation}

Finally, we set the constraint in Equation~\ref{eq:constraint_capacity} that the sum of the parallelism for the different operators should be exactly the number of task slots $\mathcal{P}$: 

\vspace{-2mm}
\begin{equation}
    \sum_{i}\pi^i = \mathcal{P}.
    \label{eq:constraint_capacity}
\end{equation}

Since we have both discrete ($\pi_i$) and continuous ($\lambda_{\mathrm{src}}$ and $\lambda_i$) decision variables coupled to linear constraints and objective function, our problem belongs to the Mixed Integer Linear Programming (MILP) class.
Problems of this class are usually NP-complete, which is not an issue in practice for our instances' sizes.
The CO can solve this model with any classic MILP solver using branch-and-bound based algorithms.
The general behavior of solvers is to return the first optimum they reach, even if several optimums exist.
This is not an issue in our case as solutions yielding the same theoretical $\lambda_{\mathrm{src}}$ are equivalent for our purpose.

The configuration resulting from the optimization is given as an input to a second call to the CE, allowing it to determine its MST and to return it together with observed metrics. %

\section{Resource Explorer}
\label{sec:resource_explorer}

The goal of the Resource Explorer (RE) is to build the capacity planning model for a target unknown query.
It drives the collection of capacity measurements and configuration information for different resource budgets (number of tasks) and resource profiles (RAM per task).
Based on configurations and rates received from the CO, the RE can further determine an optimal configuration for a (predicted) resource budget.

The RE models the relation between resources and capacity as a \emph{surrogate model}, i.e., an approximation of a complex system replacing expensive simulation models~\cite{alizadeh2020managing}.
The RE identifies the most representative function that describes the relation between a number of TS $\mathit{\Pi}$ (all used, i.e., $\sum{\pi^i} = \Pi$) using resource profiles with $\mathit{M}$ MB of memory per TS and the resulting capacity $\lambda_{src}$ as detailed in equation~\ref{eq:prediction_general}.%

\begin{equation}
f(\mathit{M}, \mathit{\Pi}) = \lambda_{src}
\label{eq:prediction_general}
\end{equation}

While the performance of jobs containing only stateless operators can generally be represented linearly, this is not the case for jobs containing stateful operators, aggregates, and especially joins.
Based on our observations, we consider two additional monotonically increasing functions with a derivative that decreases slowly over time, logarithmic and square root functions.
The three linear regressions are given in equations~\ref{eq:lin},~\ref{eq:log}, and~\ref{eq:sqrt}, with $a$ and $b$ the slopes of the functions of the surrogate models, applied to the quantity of memory and task slots, and $c$ the intercept. 
The goal of the RE is to determine which of the three models fits best the observations, find coefficients $a$, $b$, and $c$, and use the chosen model.

\vspace{-5mm}
\begin{gather} %
  a \mathit{M} + b \mathit{\Pi} + c = \lambda_{src} \label{eq:lin}\\
  a  \log \mathit{M}  + b  \log \mathit{\Pi} + c = \lambda_{src} \label{eq:log}\\
  a  \sqrt{\mathit{M}} + b \sqrt{\mathit{\Pi}} + c = \lambda_{src} \label{eq:sqrt} 
\end{gather}
\vspace{-4mm}

\smallskip
\noindent
\textbf{Overview.}
The RE builds concurrently the three candidate models based on a succession of measurements by the CO.
The model is then used to extrapolate, for high values of $\lambda_{src}$, values of $\mathit{\Pi}$ that are outside of that search space.

The construction of surrogate model candidates must balance their extrapolation capacity and the cost of collecting measurements.
The collection of the MST for a given resources budget and resource profile requires running the CO and up to two instances of the CE.
Due to the need for the query to stabilize to its steady state, these steps easily represent dozens of minutes of data collection in the test cluster.

The model construction happens in three phases.
First, in a \emph{candidate search} phase, we use Bayesian Optimization (BO) to determine new resource budgets and profile candidates optimizing the distance between the prediction of the current best model and the actual results of the runs.
Second, in a \emph{model selection} phase, we determine which of the three models has the best potential for extrapolation.
We evaluate this potential for each model on the half of the measurement set with the highest number of TS, based on the other half.
Finally, as the best model is selected, it can be used to directly predict the MST for a given configuration.
For capacity planning, a \emph{model usage} phase reverse solves the model and uses busyness and processing rate metrics collected from the CO to compute an appropriate configuration.

\smallskip
\noindent
\textbf{Model performance.}
We measure the accuracy of the candidate models with the root mean squared error (RMSE), a quantitative measure of how well the model predicts the resource requirements by comparing predicted values with the results obtained by the CO.
Lower RMSE values indicate a better-performing model.

The predictive capabilities of the models are determined using Leave-One-Out Cross-Validation (LOOCV): for each observation in the data set, we evaluate if the model trained on all other observations provides a good (low) RMSE value.
LOOCV helps to minimize overfitting and provides a reliable estimate of the model's predictive capabilities, especially in cases where there are few observations~\cite{ng1997preventing}.

\smallskip
\noindent
\textbf{Candidate search.}
The goal of the candidate search is to find a number of candidate couples ($\mathit{M}$, $\mathit{\Pi}$) and their corresponding MST that reduce the training error for the current best model.

The search space is 2-dimensional.
The number of task slots $\mathit{\Pi}$ ranges from the number of operators of the query to the number of cores in the test cluster.
$M$ is discretized using a level of granularity. %
The maximal value for $M$ is the largest possible in the test cluster when divided by all cores.

If we consider a query with 9 operators on a test cluster with 48 cores (39 possible values for $\mathit{\Pi}$) and memory ranging from 512~MB to 4~GB in increments of 512~MB (8 possible values) then the search space admits 312 different combinations.

An intensive evaluation of the search space (i.e., a grid search~\cite{dufour2019finite}) is simply too costly. 
We use instead Bayesian Optimization (BO)~\cite{antonio2021sequential}, an optimization method targeting black-box functions, often used for expensive-to-evaluate problems such as hyper-parameter optimization~\cite{feurer2019hyperparameter}.
BO is based on a \textit{probabilistic model} (a Gaussian Process) approximating the true function coupled to an \textit{acquisition function} guiding the search for the optimal point regarding a \textit{cost function} supplied by the user. 
The acquisition function balances exploration (searching in uncertain locations of the search space) and exploitation (searching in regions with low predicted error) of candidate points, collected in set $D$.
Typical acquisition functions include Probability of Improvement, Expected Improvement, and Lower Confidence Bound.
The RE uses Expected Improvement as it balances the search for low mean (exploitation) and high variance (exploration) candidates.

The candidate search must identify configurations that are likely to yield better RMSE values for the currently better candidate model.
The cost function in equation~\ref{eq:compute_error} identifies the lowest LOOCV score amongst candidate models, reducing the training error.
The general behavior of the training process is that, following an initial chaotic phase, the RMSE decreases until new individuals start to worsen the score.
Our stop criterion takes this behavior into account.
We proceed to a minimum of 3 measurements in addition to the 4 corners used to bootstrap $D$.
Then, we stop when the RMSE increases by more than 10\% between two measurements, or when reaching the maximal number of 20 measurements (by default).

\vspace{-3mm}
\begin{equation}
    \text{BestModel}(D) = \argmin_{\text{model} \in \{\text{linear}, \text{log}, \text{sqrt}\}} \text{LOOCV}(D, \text{model})
    \label{eq:compute_error}
\end{equation}

The set $D$ is bootstrapped by collecting measurements from the 4 ``corners'' of the search space, i.e., 4 combinations of lowest and highest values of $\mathit{\Pi}$ and $\mathit{M}$, enabling to compute an initial LOOCV score for the three candidate models.

The BO process then uses the candidate search by calling the acquisition function, a call to the CO followed by the computation of the error using Equation~\ref{eq:compute_error} as the cost function, and then the adaptation of the internal probabilistic models based on the received candidate couple ($\mathit{M}$, $\mathit{\Pi}$).
Note that, to account for the unavoidable measurement variations that happen when collecting MST from the CO and the CE, the RE may decide to evaluate anew a candidate couple that has been previously evaluated, to reduce uncertainty.
Our experience is that, indeed, variations are common between two runs with the same budget and profile in particular for complex queries involving joins and/or windowed operations.

\smallskip
\noindent
\textbf{Model selection.}
Following the BO, we select the most appropriate model from the three candidates based on their predictive capability.
The first half of $D$ with observations with the lowest values of $\Pi$ serves as the training set.
The other half serves as the test set. %
We apply each candidate model and evaluate how well it predicts observations in the test set.
The model with the lowest average RMSE for points in the test set is selected (obviously, we use the model trained on the full $D$ and not only on the training set used for model selection).

\smallskip
\noindent
\textbf{Model usage.}
The model considers the resources budget and their profile as independent variables.
It can be used directly to derive the capacity.
Returning the necessary resources budget for a target input rate requires, however, to solve the model inversely.
We adopt an iterative strategy where we consider one or several memory value(s) $M$\footnote{The user's production cluster may already be configured with a specific profile, or she may want to know which profile is best for her query.} and we incrementally increase $\Pi$ until the predicted capacity matches or exceeds the request.

\sysname tends to estimate resource budgets for which the requested rate is very close to the max capacity.
To guarantee that the requested rate will be sustainable, the RE applies a slight over-provisioning factor by interrogating the model with 110\% (by default) of the requested rate.

The model returns the number of task slots and their profile but not their configuration.
If requested, this configuration can be computed using a final pass of BIDS2 using the true processing rates and busyness levels collected for the largest number of cores in $D$ with the selected resource profile(s).

\section{Implementation}
\label{sec:implementation}

\sysname uses a cloud-native stack with Kubernetes~\cite{k8s} supporting a data lake using Apache Kafka~\cite{kreps2011kafka} for data ingestion, storage, and replay and \flink for processing.
We use the Strimzi Kafka Kubernetes operator~\cite{strimzi} to use Kafka as a source of data for Flink, and Spotify's \flink Kubernetes operator~\cite{spotify_k8s} to be able to deploy \flink easily with different memory and CPU settings (Kubernetes operators support specific software in Kubernetes and are not \flink operators).

\smallskip
\noindent
\textbf{Capacity Estimator.}
The CE interacts with a running, unmodified instance of \flink v.1.14.1~\cite{flink_1.14.1}, using Apache Zeppelin~\cite{zeppelin} notebooks.
Runtime measurements (e.g., about processing rates and busyness) are collected using Prometheus~\cite{prometheus} with a 5-second aggregation period.

We \flink instance in the test cluster without auto-scaling features, using only dedicated cores for every TS without simultaneous multi-threading. %
This \flink instance is deployed by the CE on demand with a fixed, homogeneous resource profile for the TS, and reused if possible.
If the CO requests a different profile, the necessary redeployment takes less than a minute.

The CE must be able to control precisely the rate at which the source of a query under test in the small cluster receives data.
We developed a novel \emph{rate-limited} Kafka source connector extending the regular Apache Flink Kafka source~\cite{kafka_connector}.
This rate-limited source obtains data from a first Kafka topic as well as rate control information from a secondary topic.

\smallskip
\noindent
\textbf{Configuration Optimizer and Resource Explorer.}
The RE and CO are both developed in Python.
The CO uses the Python PuLP library \cite{pulp} coupled to the CBC (Coin-or Branch and Cut) solver~\cite{CBC} both developed by the COIN-OR foundation. 
Metrics are retrieved from CE runs in Prometheus using the prometheus-api-client library~\cite{prometheus-api-client}.
The RE uses the scikit-optimize~\cite{scikit-optimize} library for Bayesian Optimization and the scikit-learn~\cite{scikit-learn} library for the regression.

\section{Evaluation}
\label{sec:evaluation}

We evaluate \sysname in a large cluster of 85 nodes using large-scale representative queries.
We wish to answer the following research questions.
  (1)~Is \sysname able to predict, based on small-scale runs, an accurate budget of resources, their profile, and their configuration, such that a production run based on this prediction is neither over- or under-provisioned?
  (2)~How much resources and time are necessary for \sysname to build capacity planning models?
To answer these high-level questions, we first answer the following.
  (3)~Is the evaluation of the MST by the CO and CE accurate, i.e., can the RE rely on the measurements for its model training?
  (4)~Is the CO effective at deriving the ``best possible'' configuration for a given resource budget?

We first present our workloads and our experimental setup.
Then, we answer questions~(3) and~(4) using micro-benchmarks at the CE and CO levels.
Finally, we answer questions~(1) and~(2) using macro-benchmarks involving the RE and the full \sysname stack as well as a comparison to production-scale deployments.

\begin{figure}[t]
    \centering
    \includegraphics[scale=\omnifiguresscalingfactor]{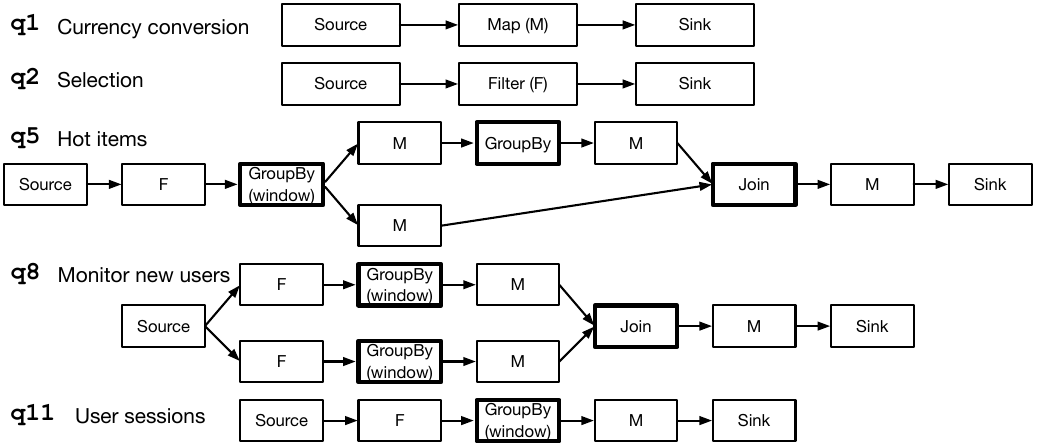}
    \vspace{-1mm}
    \caption{Operator graphs for Nexmark queries.
    Stateful operators are with a bold frame.
    Stateless operators are either map (transform input items) or filter (transmit a subset of the input).
    \vspace{-5mm}
    }
    \label{fig:queries}
\end{figure}

\smallskip
\noindent
\textbf{Target workload and queries.}
We use representative queries from the \flink SQL implementation of  Nexmark~\cite{tucker2008nexmark,nexmark_website}, a reference benchmark also used in the evaluation of DS2~\cite{kalavri2018three}.
The benchmark reproduces an online auction system, with various continuous queries.
According to our target assumptions, we consider queries with no continuously-increasing state.
The queries are summarized in Table~\ref{tab:queries} and represented in Figure~\ref{fig:queries}. 
Nexmark provides a data generator that we use with its default settings, i.e., the input event stream features 2\% of events linked to persons, 6\% proposing auctions, and 92\% representing bids by the former to the latter. 
The average size of these events are respectively 200, 500, and 100~Bytes.
We populate the data lake (Kafka) with pre-generated data streams that we use as data-at-rest input for \sysname operations.

\begin{table}[t]
\begin{center}

\newcolumntype{G}{>{\rowfonttype\centering\let\newline\\\arraybackslash\hspace{0pt}}m{0.1\linewidth} }
\newcolumntype{H}{>{\rowfonttype\centering\arraybackslash} m{0.4\linewidth} }
\newcolumntype{I}{>{\rowfonttype\centering\arraybackslash} m{0.125\linewidth} }
\newcolumntype{J}{>{\rowfonttype\centering\arraybackslash} m{0.2\linewidth} }
\newcolumntype{K}{>{\rowfonttype\centering\arraybackslash} m{0.18\linewidth} }

\renewcommand{\arraystretch}{1.1}
\setlength{\tabcolsep}{0pt}

\begin{tabular}{ G H I J K }
  \toprule
  \rowfont{\scriptsize}
    \bf Query &
    \bf Description &
    \bf Operators &
    \bf Stateful &
    \bf Min rate ($\times 10^3$ evt/s) \\
\midrule  
  \texttt{q1}  & \textbf{Currency conversion}: converts bid values from dollars to euros 
    & 1 & - & 1600 \\
  \texttt{q2}  & \textbf{Selection}: filter bids with specific auction identifier 
    & 1 & - & 3600 \\
  \texttt{q5}  & \textbf{Hot items}: determine auctions with most bids in last period
    & 8 & GB, GBW, J & 50 \\
  \texttt{q8}  & \textbf{Monitor new users}: identify active users in last period
    & 8 & GBW ($\times$2), J & 1400 \\
  \texttt{q11} & \textbf{User sessions}: compute number of bids each user makes while active
    & 3 & GBW & 60 \\
  \bottomrule
\end{tabular}%
\end{center}
\caption{Nexmark queries~\cite{tucker2008nexmark,nexmark_website}.
The number of operators excludes sources and sinks.
``Stateful'' lists stateful operators used by the query: GB for GroupBy, GBW for GroupBy (window), and J for Join.
Minimal rates are for single-task configurations with 4-GB profiles.
\vspace{-2mm}
}
\label{tab:queries}
\end{table}

Queries \texttt{q1} and \texttt{q2} use a single, stateless operator.
The other queries include stateful operators including GroupBy (window) and Joins.
Queries \texttt{q5} and \texttt{q8} have complex graphs with 8 operators, including GroupBy and Joins.
Query \texttt{q11} uses a pipeline of three operators, including a compute-heavy GroupBy (window).
All stateful functions use a time window of 10 seconds.
For \texttt{q5}, windows slide in increments of 2 seconds while they are non-overlapping for \texttt{q8} and \texttt{q11}.

\begin{figure*}[t!]
  \centering
  \includegraphics[scale=\fpeval{\omnifiguresscalingfactor}]{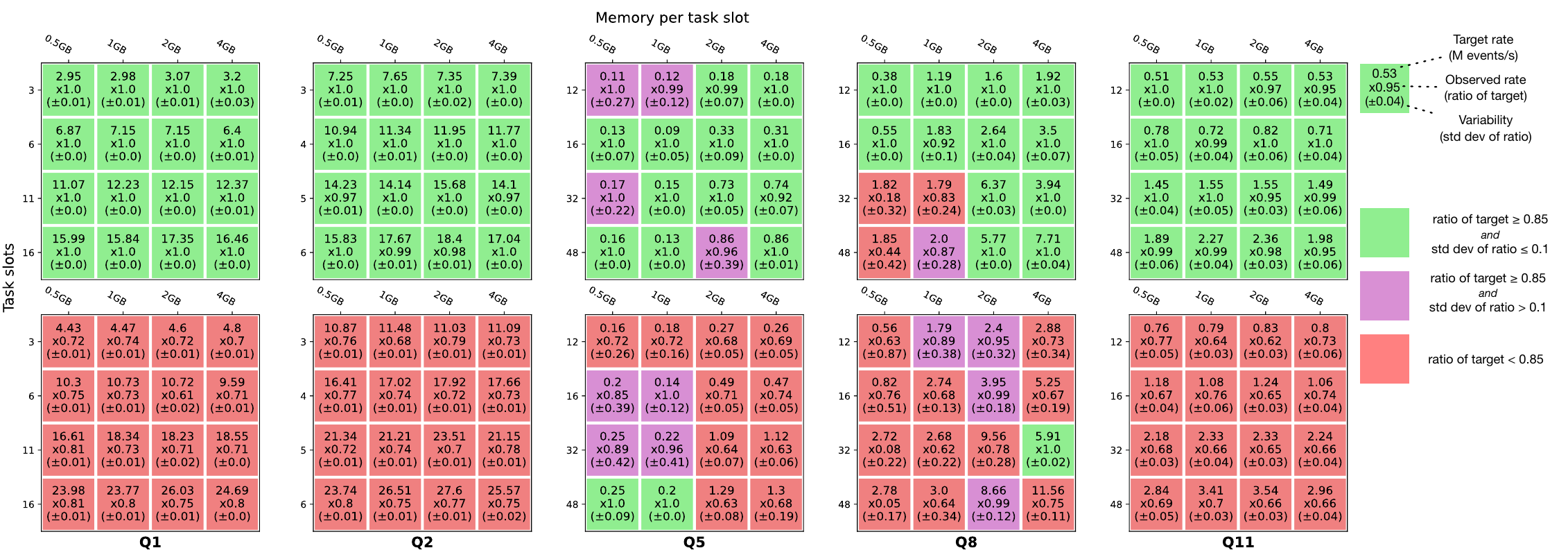}
  \vspace{-2mm}
  \caption{Accuracy of the MST estimation under various resource budgets and task profiles.
  The MST estimated by the CE for the configuration decided by the CO is replayed at 100\% for the upper row and at 150\% for the lower row.
  Colors represent runs that sustainably process the target rate (green), approach or meet the target with instabilities (purple), or fail (red).
  \vspace{-3mm}
  }
  \label{fig:xp_r1b}
\end{figure*}

\smallskip
\noindent
\textbf{Experimental setup.}
Each node in our cluster has an 18-core Intel Xeon Gold 5220 and 96~GB of RAM.
Their 480-GB SSD has sequential read and write performance of 540 and 520~MB/s.
Nodes are connected with 25-Gbps Ethernet.

\sysname needs sufficiently many nodes supporting data injection and source operators (if these are under-dimensioned, \sysname may return unsuccessfully after failing to replay a target rate).
Table~\ref{tab:queries} presents as a reference the minimal rate that a single-task configuration supports for each query.
Queries \texttt{q1}, \texttt{q2}, and to a lesser extent \texttt{q8} have a high minimal rate, while \texttt{q5} and \texttt{q11} process fewer events per second.

When replaying data at rest, the limiting factor are our modest SSDs, with a peak capacity per Kafka server of about 1,200,000 events/s.
This represents, for our target queries, a minimal number of 8 Kafka nodes for $q5$ and $q11$ and 16 Kafka nodes for $q1$, $q2$, and $q8$ to be able to inject rates decided during the RE exploration (obviously, using machines with multiple, high-performance SSD or HDD would allow reducing this number).
Rate-limited source operators are CPU-bound. 
We must also ensure they never are the limiting factor when testing a specific configuration.
We identified that 64 source tasks are necessary to support the maximal rate that the Kafka nodes can serve (4 servers).

The test cluster uses 3 servers with 48 task slots to leave 2 available cores per machine for system management.
Similarly, we use up to 64~GB of RAM per machine, resulting in a maximum profile of 4~GB per task.
We also dedicate 3 nodes for Kubernetes, Grafana, and the \flink Job Manager.%

\smallskip
\noindent
\textbf{MST estimations.}
We first evaluate the accuracy of the CO and CE in identifying the MST of small-scale runs.

A run for the CE with queries \texttt{q1}, \texttt{q2}, and \texttt{q11} uses a warmup of 120 seconds and measurements of 75 seconds (60 seconds of local ramp-up and 30 seconds of observation).
The cooldown of 15 seconds uses a throughput of 200 events per second and per source, for a total of 6,400 events/s.
One CE run performs 8 iterations for a total duration of 645 seconds (10.75 minutes).
For complex queries \texttt{q5} and \texttt{q8}, our initial evaluations highlighted the need for longer measurements, due to the longer time necessary for our configurations to use RocksDB disk storage and not only the in-memory cache: We use a 450-second warm up, a 900-second duration, and 7 iterations, for a total of 900 seconds (15 minutes); we also increase cooldown throughput to 12,800 events/s.

We query the CO for 16 combinations of resource budgets and profiles per query.
We consider 3 to 16 tasks for \texttt{q1}, 3 to 6 for \texttt{q2}, and 12 to 48 for \texttt{q5}, \texttt{q8}, and \texttt{q11} (the high achievable rate of \texttt{q1} and \texttt{q2} makes it unnecessary to test them with the full cluster).%
We use memory profiles of 0.5, 1, 2, and 4~GB.

We run each query with the 16 configurations returned by the CE, using their estimated MSTs as the target rate.
We observe if each configuration supports the estimated rate sustainably for 10 minutes preceded by a warmup of 2 minutes.%
Figure~\ref{fig:xp_r1b} presents the results in its upper row.
For each of the 16 configurations, a box presents first the target rate returned by the CE for this configuration (for instance, \texttt{q8} with 12 TS of 2~GB has an MST of 1.92$\times 10^6$ events/s) and the observed rate, expressed as a ratio of this target.
A high level of variation in the measurements, as presented by the standard deviation of this ratio across all measurements, is a sign that the job is close or past saturation.

Finally, we report in the lower row of Figure~\ref{fig:xp_r1b} results using a target rate of 150\% of the estimated MST.
A configuration passing this test means that its estimation was too conservative.

\begin{figure*}[t!]
  \centering
  \includegraphics[scale=\fpeval{\omnifiguresscalingfactor * 0.95}]{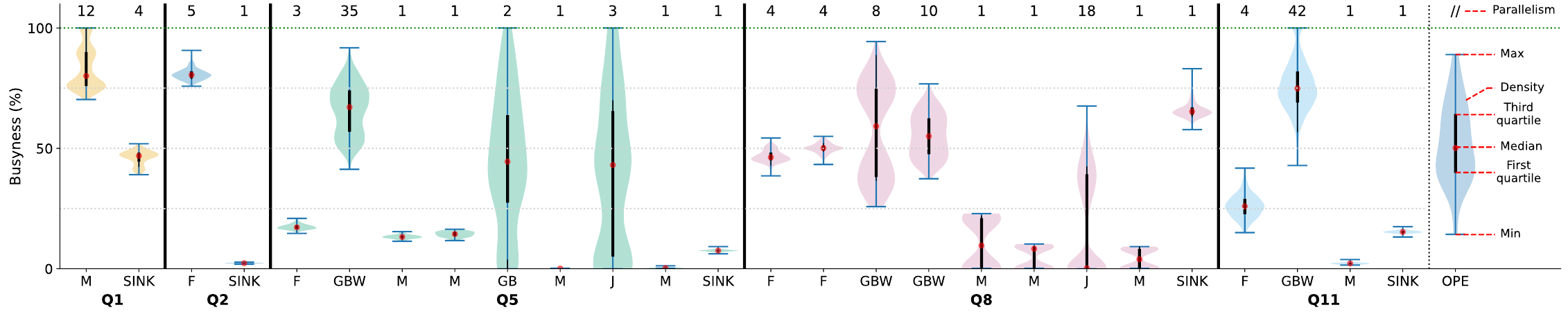}
  \vspace{-1mm}
  \caption{Distribution of task busyness levels measurements.
  Numbers above box-plots are the parallelism of operators.
  \vspace{-3mm}
  }
  \label{fig:xp_r3}  
\end{figure*}

Simple queries \texttt{q1} and \texttt{q2} show little variations.
All their tested configurations sustainably process the estimated MST at 100\% but fail with 150\%.
Query \texttt{q11} shows good results but slightly higher variations.
This is due to the behavior of its stateful GroupBy (window) operator and resulting stragglers.
For these three queries, a general linear scale behavior seems to apply with increasing numbers of TS.
The impact of memory is less clear and highlights the variability of the in-situ data that the RE receives as an input (and that the BO will address by repeating runs where the RMSE error is important).

For complex queries \texttt{q5} and \texttt{q8}, we observe a lower level of repeatability, and low to scaling capabilities in terms of task slots.
For \texttt{q5}, our investigation is that the factor of instability lies mostly in the Join operator and its very uneven load across tasks, as we also highlight in the next experiment.
For \texttt{q8}, variability is caused by the presence of stragglers due to the windowed operators using a non-overlapping window size of 10 seconds, lower than our 5-second monitoring period, and leading to ``sawtooth-like'' load profiles.
For both queries, results with 150\% of the MST injected show that some estimations (in particular with 32~TS/4~GB and 48~TS/2~GB for \texttt{q8}) can be too conservative.
This behavior is not predictable from one run to the next, and we did not select a ``good run'' but the one performed in the whole series.
This indicates again that the RE must accommodate variations in the data collected in particular for more complex queries.

For all memory profiles, \texttt{q1}, \texttt{q2}, and \texttt{q11} achieve at worst 95\% of the expected rate, and often 99-100\%.
For \texttt{q5} and \texttt{q8}, performance is much lower and very volatile with 0.5 and 1~GB, highlighting the impact of insufficient memory.
We focus on 2- and 4-GB profiles for these two queries.

\smallskip
\noindent
\textbf{Configurations.}
We now evaluate the result of the CO optimization, zooming in the largest configuration for each query running at 100\% of the estimated MST (i.e., bottom-right corners of matrices in Figure~\ref{fig:xp_r1b}'s first row).
Figure~\ref{fig:xp_r3} presents the distribution of the time series measurements over each 10-minute run.
Most scaled-out operators reach their peak capacity at some point in time during the run, even if the median busyness is lower.
This is the expected behavior: 
  Operators must be provisioned to handle peak load.
For Group By (window) and Joins, we observe a wide range of busyness levels, due to the skew and the stragglers resulting from upstream Group By (window) operators' uneven output rates.
For the Join of \texttt{q8}, additional tests with a forced lower parallelism (not shown) lead to saturations, back-pressure cascades, and instabilities; the measured busyness of 60\% seems to be a practical maximum.
Some stateless, scaled-out operators such as the first Filter of \texttt{q5}, \texttt{q8}, and \texttt{q11} have relatively low busyness levels, due to an important level of back-pressure from downstream tasks.
Overall, the CO can avoid under-provisioned operators with frequent busyness levels of 100\%.

\smallskip
\noindent
\textbf{Capacity planning.}
Our final evaluation explores the capacity of \sysname to build an effective and accurate capacity planning model and the cost of building this model.

Table~\ref{tab:model_parameters} presents the results of a run of the RE for each of the five queries, with the number of calls to the CO and CE, the training time, and the resulting models.
Queries \texttt{q1}, \texttt{q2}, and \texttt{q11} are identified with a linear scaling model, while \texttt{q5} and \texttt{q8} respectively use the log and square root models.

The training by the RE uses 9 (for \texttt{q1}) to 16 (for \texttt{q11}) calls to the CO.
These calls result in 10 to 20 calls to the CE, as not all CO calls require evaluating a single-task configuration that is already in the cache.
The complete duration of a CO call is about 24-minute long when the single-task configuration did not run, and about 13-minute when it did (for \texttt{q5} and \texttt{q8} a CO call takes 18 minutes always, as single-task configurations are tested first as part of the corners to bootstrap $D$).
The CO optimization itself takes about 100~ms, while BO steps at the RE level have a negligible duration ($<$1~ms).
In total, the total training duration ranges from 139~minutes for \texttt{q1} to 252~minutes for \texttt{q5}.
These durations are reasonable considering the costs and scale of the target production deployments.

\begin{table}[t]
    \begin{center}
    
    \newcolumntype{A}{>{\centering\arraybackslash} m{0.08\linewidth} }
    \newcolumntype{B}{>{\centering\arraybackslash} m{0.1\linewidth} }
    \newcolumntype{C}{>{\centering\arraybackslash} m{0.1\linewidth} }
    \newcolumntype{D}{>{\centering\arraybackslash} m{0.08\linewidth} }
    \newcolumntype{E}{>{\centering\arraybackslash} m{0.15\linewidth} }
    \newcolumntype{F}{>{\centering\arraybackslash} m{0.1\linewidth} }
    \newcolumntype{G}{>{\centering\arraybackslash} m{0.1\linewidth} }
    \newcolumntype{H}{>{\centering\arraybackslash} m{0.1\linewidth} }
    \newcolumntype{I}{>{\centering\arraybackslash} m{0.1\linewidth} }
    \newcolumntype{J}{>{\centering\arraybackslash} m{0.1\linewidth} }
    
    \renewcommand{\arraystretch}{1.1}
    \setlength{\tabcolsep}{0pt}
    
    \begin{tabular}{ A B C D D E F G H I}
      \toprule
      \rowfont{\scriptsize}
        & 
        \multicolumn{2}{c}{\textbf{-- Min/Max --}} &
        \multicolumn{2}{c}{\textbf{-- Runs --}}&
        &
        &
        \multicolumn{3}{c}{\textbf{---- Coefficients ----}}
         \\      
        \bf Query &
        \bf TS &
        \bf RAM &
        \bf \#CO &
        \bf \#CE &
        \bf Duration &
        \bf Model &
        \bf $a$ &
        \bf $b$ &
        \bf $c$ \\
    \midrule  
      \texttt{q1}  & 2/16 & 0.5/4 & 9 &  10 & 139~min. & lin  & 1.0 & 9.9E5 & -7.6E5 \\
      \texttt{q2}  & 2/6 & 0.5/4  & 14 & 20 & 248~min. & lin  & 7.5 & 3.0E6 & -2.7E6 \\
      \texttt{q5}  & 9/48 & 2/4   & 14 & 14 & 252~min. & log  & -7.6E3 & 5.7E5 & -1.2E6 \\
      \texttt{q8}  & 9/32 & 2/4   & 13 & 13 & 234~min. & sqrt & 2.6E3 & 1.4E6 & -3.9E6 \\
      \texttt{q11} & 4/48 & 0.5/4 & 16 & 20 & 252~min. & lin  & 4.1 & 3.9E4 & -2.1E5 \\

      \bottomrule
    \end{tabular}
    \end{center}
    \caption{RE training results and costs, chosen model, and coefficients. \#CO/\#CE are number of calls to the two modules.}
    \label{tab:model_parameters}
\end{table}

\begin{table}[t]
  \begin{center}
    
    \newcolumntype{G}{>{\centering\arraybackslash} m{0.15\linewidth} }
    \newcolumntype{H}{>{\centering\arraybackslash} m{0.3\linewidth} }
    \newcolumntype{I}{>{\centering\arraybackslash} m{0.1\linewidth} }
    
    \renewcommand{\arraystretch}{1.1}
    \setlength{\tabcolsep}{0pt}
    
    \begin{tabular}{ G H I I I I }
      \toprule
      \rowfont{\scriptsize}
       &
       &
       \multicolumn{4}{c}{\bf Number of TS with profile} \\
      \bf Query &
      \bf Requested rate &
      \bf 0.5~GB &
      \bf 1~GB &
      \bf 2~GB &
      \bf 4~GB \\
      \midrule  
      \texttt{q1}  & 160 $\times 10^6$ & 179 & 179 & 179 & 178 \\
      \texttt{q2}  & 190 $\times 10^6$& 69 & 69 & 69 & 69 \\
      \texttt{q5}  &  2.5 $\times 10^6$ & - & - & 1069 & 1079 \\
      \texttt{q8} & 15 $\times 10^6$ & - & - & 179 & 176 \\
      \texttt{q11}  & 20 $\times 10^6$ & 565 & 564 & 562 & 559 \\
      \bottomrule
    \end{tabular}
  \end{center}
  \caption{Capacity planning results for the five queries.}
  \label{tab:model_results}
\end{table}

Table~\ref{tab:model_results} presents the uses of the models for large requested rates.\footnote{Querying for the resources meeting a requested rate requires solving the inverse problem, which takes at most a few seconds.}
We can see that the models identify a minor impact of memory profiles for most queries except \texttt{q11} (also clear in the values of coefficient $a$ in the models).
For \texttt{q5}, the small variation ($<$1\%) between the 2- and 4-GB profiles is a result of the uncertainty in the CO measurements.
We can also observe that the effect of increasing rates differs between tasks, e.g., with coefficient $b$ in the linear models of \texttt{q1}, \texttt{q2}, and \texttt{q11}.

We validate the predictions using large-scale production runs.
These runs use up to 85 nodes, using up to 69 nodes for \flink TMs (for \texttt{q5}) and up to 36 Kafka nodes (for \texttt{q11}).
As our cluster does not allow supporting with Kafka the rates requested for \texttt{q1} and \texttt{q2} we inject for these queries data directly using sources paired with the Nexmark generator.

We adopt a similar strategy as for testing small-scale runs to verify that the predictions of \sysname are neither over- or under-provisioned.
A prediction is not under-provisioned if it sustainably supports the requested rate over time.
A prediction is not over-provisioned if, when injecting a higher rate (we use 120\% and 150\% of the requested rate) the query shows signs of instability or insufficient capacity.
Instability can manifest in the observed, actual rate that we expect to match with no variation in the requested rate.
We also observe the \emph{pending records} metrics, i.e., how many events generated by the fixed-rate source ``pile up'' at the source operator due to back pressure.
While a small amount of buffering is normal in a setup close to full utilization, an ever-increasing pending records metric is a clear sign of an under-provisioned system.

\begin{figure}[t]
  \textbf{\texttt{q1~}}
  \includegraphics[scale=\fpeval{0.95*\omnifiguresscalingfactor}]{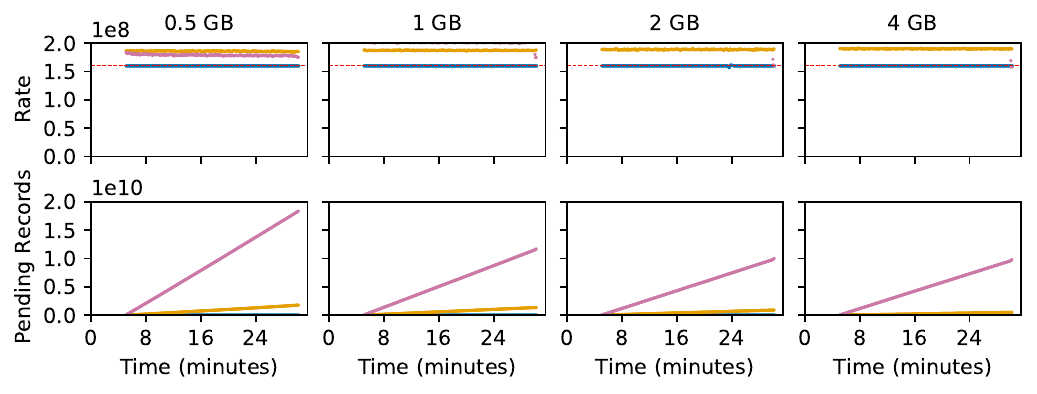} \\
  \textbf{\texttt{q2~}}
  \includegraphics[scale=\fpeval{0.95*\omnifiguresscalingfactor}]{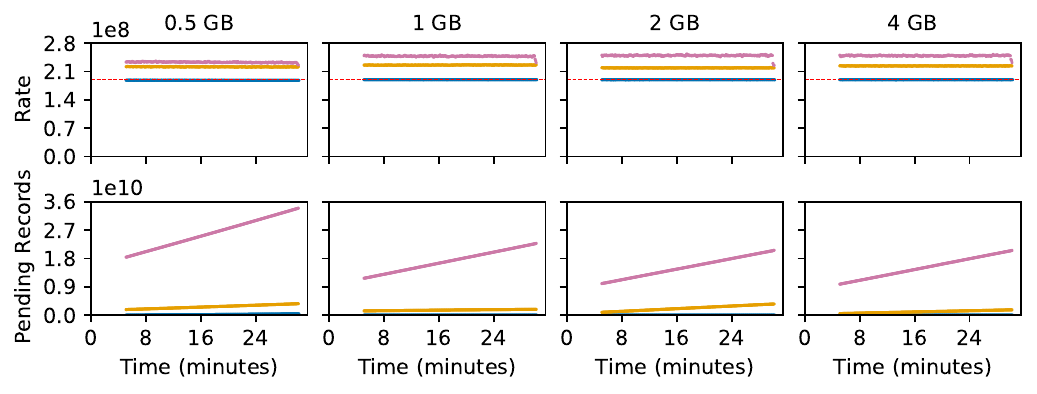} \\
  \textbf{\texttt{q5~}}
  \includegraphics[scale=\fpeval{0.95*\omnifiguresscalingfactor}]{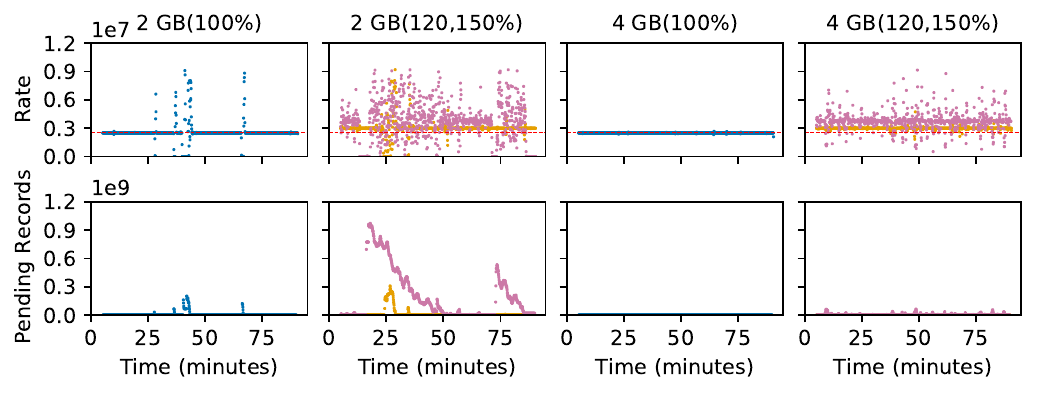} \\
  \textbf{\texttt{q8~}}
  \includegraphics[scale=\fpeval{0.95*\omnifiguresscalingfactor}]{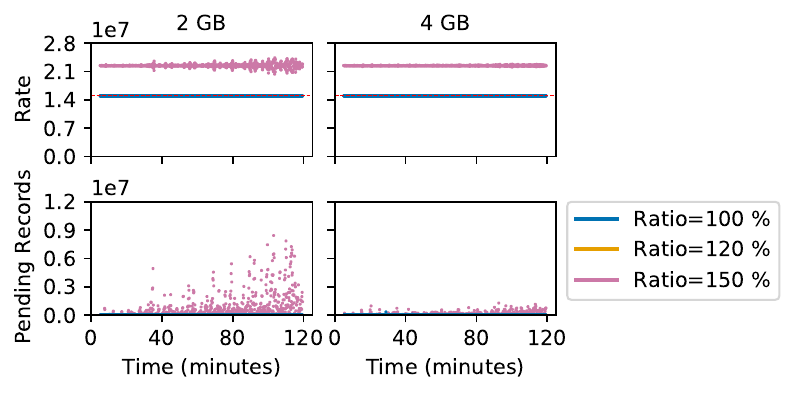} \\
  \textbf{\texttt{q11}}
  \includegraphics[scale=\fpeval{0.95*\omnifiguresscalingfactor}]{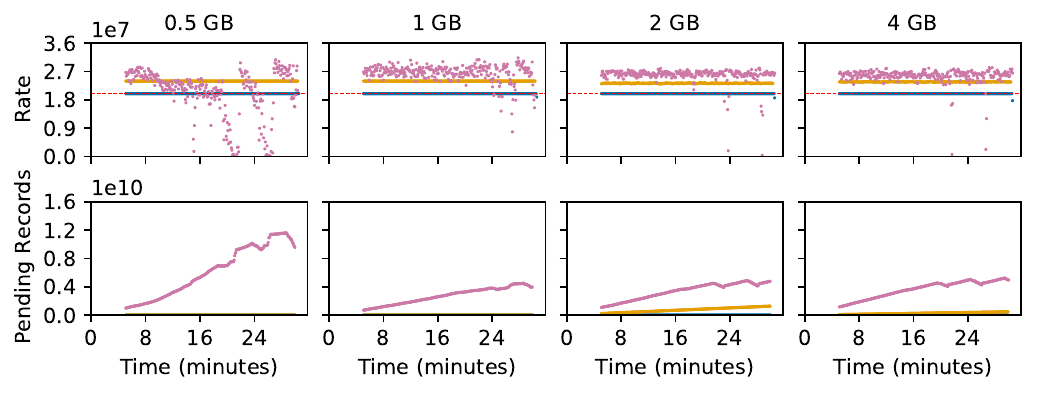}
  \caption{Large-scale production runs using capacity planning models (Table~\ref{tab:model_results}).
  For each query, we present the requested rate (thin red line) and the actual rate when injecting 100\% (blue), 120\% (yellow), and 150\% (purple) of this requested rate.
  The second row present accumulated pending records.
  We present the 100\% and (120\%, 150\%) cases separately for each of the two profiles of \texttt{q5} to improve readability.
  }
  \label{fig:xp:macrobenchmark}
\end{figure}

Figure~\ref{fig:xp:macrobenchmark} presents the results of production-scale runs using the number of TS (cores) given in Table~\ref{tab:model_results}.
We measure rate and pending records metrics after a ramp-up period of 5 minutes, for a duration of 30 minutes (\texttt{q1}, \texttt{q2}, and \texttt{q11}), 90 minutes (\texttt{q5}), and 120 minutes (\texttt{q8}).
For \texttt{q1} and \texttt{q2}, the actual rate is exactly the one requested at 100\%, while 120\% and 150\% rates are not matched and lead to ever-increasing pending records.
We note that after 30 minutes, \texttt{q2} with the 0.5~GB profile shows a slight delay in processing events (equivalent of 1.6~s worth of pending records), illustrating the difficulty of running even stateless queries at large-scale and high-throughput with \flink under memory constraints (primarily due to the JVM garbage collections).
We do not observe this behavior for other setups, including stateful \texttt{q11} with all profiles.

For \texttt{q5}, for readability, we present separately the 100\% rate and the two other rates for the two considered profiles (2~GB and 4~GB).
The 100\% rate is sustained in both cases but with 2~GB per TS, we observe temporary instabilities.
Resulting pending records are, however, later absorbed by the job indicating that the saturation point is not reached.
With 4~GB the query is very stable.
In both cases, we observe chaotic behavior in terms of throughput for 120\% and 150\%, meaning the saturation point was reached.

For \texttt{q8} we consider only the 100\% and 150\% cases for readability.
Interestingly, we can see the impact of long-term instabilities that \sysname can account for in the model (i.e., as these show earlier on small-scale jobs than on large-scale ones with high parallelisms for the concerned operators).
Here, while the query is initially able to process 150\% of the requested rate for a time, we observe that pending records gradually accumulate and provoke instabilities, making the query non-sustainable. 
The reason is that the working set of the query fits for a time dependent on the total memory before introducing congestion.
We observe this effect both with 2- and 4-GB profiles.

\smallskip
\noindent
\textbf{Summary.}
Our evaluation answer positively to our research questions.
We observe that the collection of individual measurements is most often accurate, but can yield noisy outputs when memory is insufficient for the considered query.
Despite the uncertainty, our production-scale evaluation shows that \sysname was able to correctly perform capacity planning, even for queries that have a non-linear scaling profile.

\section{Related work}
\label{sec:related_work}

We review work on DSP performance analysis, modeling, and prediction, and on capacity planning for other contexts. %
We omit a discussion of the large body of work on elastic scaling of DSP engines but refer to existing surveys~\cite{roger2019comprehensive,cardellini2022runtime}.

\smallskip
\noindent
\textbf{DSP Benchmarking.}
The first class of systems targets the in-situ performance evaluation of DSP using benchmarking methodologies~\cite{chintapalli2016benchmarking,lopez2016performance}.
In contrast with \sysname, these systems target the deployment of a query at a production scale and do not intend to extrapolate from small-scale deployments.
StreamBench~\cite{lu2014stream} proposes, for instance, a set of benchmark programs for Apache Storm and Apache Spark Streaming.
Theodolite~\cite{henning2021theodolite,henning2022benchmark} is a benchmarking framework for \flink and Kafka Streams.
Theodolite implements different search strategies piloting tests with varying CPU budgets, verifying if service-level objectives (SLO) are met for each of them, and forming a scalability profile.
The binary search strategy resembles the method used in the CE, although it applies to several, independent runs.
Neither works provide a solution to decide on the configuration (level of parallelism) of each operator of a query, which must be set by the user.
Rafiki~\cite{pfister2022rafiki} allows determining such configurations by piloting a series of runs and gradually increasing the parallelism of each operator based on observed backpressure metrics.
Zeuch \emph{et al.}~\cite{zeuch2019analyzing} use benchmarking to suggest that DSP engines may \emph{scale up} by making better use of modern hardware.
Gadget~\cite{asyabi2022new} is a benchmark suite targeting the performance of the storage subsystem in DSP, e.g., RocksDB and alternatives.

\smallskip
\noindent
\textbf{DSP modeling}
Another class of work proposes to \emph{model} the DSP queries, their performance, and their scalability.
Truong et al.~\cite{truong2017predicting,truong2018performance} use queueing theory to build a performance model of a specific query and its individual operators.
Complementarily, MEAD~\cite{russo2021mead} models the arrival of events at different operators using Markovian Arrival Processes to better predict performance after scale-out under bursty event patterns.

The modeling of queries can enable proactive, predictive scaling and scheduling operations~\cite{li2016performance}.
Twitter's Caladrius~\cite{kalim2019caladrius} models for this purpose the performance of parallel operators in Apache Heron using piecewise linear regression.

Some authors propose to model the performance and resource usage of queries based on general characteristics, learning from a large set of queries and identifying a general model, e.g., mixture-density networks~\cite{khoshkbarforoushha2015resource, khoshkbarforoushha2016distribution} or zero-shot cost models~\cite{heinrich2022zero,agnihotri2023zero}.
These approaches require initial training using a large set of queries (typically, several thousands).
As no dataset of \emph{real} queries of this size exists, the authors have to resort to synthetic, randomly-generated queries that may not represent real workloads.

\smallskip
\noindent
\textbf{Optimizing DSP parameters.}
Bayesian Optimization, as used in the Resource Explorer, was used to optimize the parameters of DSP engines and avoid an exhaustive sweep~\cite{fischer2015machines,jamshidi2016uncertainty, trotter2017bayesian}.
Bilal and Canini~\cite{bilal2017towards} propose, for instance, a black-box automated parameter tuning approach for Apache Storm using a hill-climbing algorithm and a rule-based, grey-box approach.

\smallskip
\noindent
\textbf{Capacity planning.}
The need to plan the configuration or scale of computing infrastructure is obviously not limited to DSP.
Higginson \emph{et al.}~\cite{higginson2020sigmod} discuss the applicability of resource forecasting techniques based on machine learning for clustered database systems.
URSA~\cite{zhang2020icpp} is a capacity planning and scheduling system for database platforms, modeling the response of a database workload to provided resources and deriving just-sufficient resource specifications automatically.
Cherkasova \emph{et al.}~\cite{cherkasova2004sla} propose a capacity planning model for media servers.
These works do not rely on controlled, small-scale testing and extrapolation as proposed by \sysname.

\section{Conclusion}
\label{sec:conclusion}

We presented \sysname, a capacity planning system for stream processing allowing to derive scaling and configuration decisions for large-scale \flink jobs from a series of controlled, small-scale runs of a target query.
Our work opens several interesting perspectives that we intend to explore in our future work.
First, we would like to consider the integration of synthetic data generation and upscaling mechanisms, as exists for relational databases~\cite{sanghi2022projection}.
Second, we observed that some queries have a clearly sub-linear scaling profile, while elastic scaling solutions generally consider linear scaling assumptions.
The models generated by \sysname, or some of its methodologies, could guide the development of more accurate elastic scalers for such queries.

\section*{Acknowledgements}
This research was funded by the Walloon region (Belgium) through the Win2Wal project GEPICIAD and by a gift from Eura Nova.

\bibliographystyle{IEEEtran}
\bibliography{bibliography}

\end{document}